\documentclass[aps,twocolumn,amsmath,amssymb,pra,graphics,graphicx]{revtex4-1}

\usepackage{graphicx}
\usepackage{bm}
\usepackage{dcolumn}
\begin{document}

\title{Anisotropic time dependent London approach}

\author{ V. G. Kogan }
\email{kogan@ameslab.gov}
\affiliation{  Ames Laboratory-DOE, Ames, Iowa 50011 }

 \author{R. Prozorov}
\email{prozorov@ameslab.gov}
 \affiliation{Ames Laboratory--DOE, Ames, IA 50011, USA}
 \affiliation{Department of Physics and Astronomy, Iowa State University, Ames, IA 50011, USA}

 \date{ \today}

\begin{abstract}
  The anisotropic  London equations taking into account the normal currents are derived and applied to  the  problem of the surface impedance in the Meisner state of anisotropic materials.  It is shown that the complex susceptibility of anisotropic slab depends on the orientation of the applied microwave field relative to the crystal axes.  In particular,  the anisotropic sample in the microwave field is a subject to a torque, unless the field is directed along one one of the crystal principle axes.
\end{abstract}

%\noindent \pacs{74.20.-z,74.20.De,74.50.+r}

\maketitle

 \section{Introduction}

Its shortcomings notwithstanding, the approach based on London equations played - and still does - a major role in describing magnetic properties of superconductors away of the critical temperature $T_c$ where it is, in fact, the only available and sufficiently simple technique  for many practical applications.
The physical reason for this success is in its ability to describe the Meissner effect, the major feature of superconductors at all temperatures.
The anisotropic version of this approach \cite{K81,BBG} has proven  useful  when strongly anisotropic high-$T_c$ materials came to the forth. It was also realized that in time dependent phenomena the normal dissipative currents due to normal excitations should be taken into account along the persistent currents \cite{CC, Blatter}. In particular, normal currents influence superconductors behavior in  microwaves absorption \cite{CC} and perturb the field distribution of moving vortices \cite{TDL,moving}.  In this work, the anisotropic version of time dependent London equation is derived and applied to problems of surface impedance and magnetic susceptibility in a simple geometry.

 Within the London approach, the current density consists, in
general, of  normal and superconducting parts:
\begin{equation}
{\bm J}= \sigma {\bm E} -\frac{c}{4\pi\lambda^2}\, \left( {\bm
A}+\frac{\phi_0}{2\pi}{\bm
\nabla}\theta\right)  \,,
\label{current}
\end{equation}
where ${\bm E}$ is the electric field, $\lambda$ is the penetration depth,  $\bm A$ is the vector potential, $\theta $ is the phase, and $\phi_0$ is the flux quantum.  The conductivity  $\sigma$     for
the quasiparticles flow is in general frequency  dependent. If however the frequencies  $\omega$ are bound by
inequality $\omega\tau_n\ll 1$ with $\tau_n$ being the scattering time for the
normal excitations, one can consider $\sigma$ as a real
$\omega$-independent quantity.  As always within the London
approach, the order parameter  is assumed constant in space.

In the absence of vortices, we have by applying  curl:
\begin{equation}
{\rm curl\,}{\rm curl}{\bm H}+\frac{1}{\lambda^2}\,{\bm H}=
-\frac{4\pi\sigma}{c^2} \frac{\partial \bm H}{\partial t} \,.
\label{London}
\end{equation}
These are in fact London equations corrected by the time dependent right-hand side \cite{TDL}.

%%%%%%
\subsubsection{Surface impedance of the half-space isotropic sample}
%%%%%%

The surface impedance in isotropic superconductors has been considered, e.g., by Clem and Coffey \cite{CC}.
Eq.\,(\ref{London}) provides a simple and direct approach to this problem. Let a weak magnetic field ${\bm H}=H_0{\hat{\bm x}}e^{-i\omega t}$ be at the
surface $z=0$ of a  superconducting  half-space $z>0$. Since the
field is assumed weak, the order parameter $f_0$ is unperturbed and we can use
the London Eq.\,(\ref{London}). The field is uniform in plane $(x,y)$ and depends only on $z$. We
look for solutions of
\begin{equation}
-\frac{\partial^2H_x}{\partial z^2}+\frac{1}{\lambda^2}\,H_x=
-\frac{4\pi\sigma}{c^2} \frac{\partial   H_x}{\partial t} \,.
\label{LondonHx}
\end{equation}
in the form $ H_x(z)\,e^{-i\omega t}$ and obtain:
\begin{equation}
H_x = H_0\,e^{-kz-i\omega t}\,,\quad k^2=
\frac{1}{\lambda^2}-\frac{2i}{\delta^2}\,,
\label{k}
\end{equation}
where $\delta=c/\sqrt{2\pi\sigma\omega}$ is the quasiparticles
related skin-depth.

The electric field is found from the Maxwell equation curl$\bm E =-\partial_t\bm H/c$: $E_y =  (i\omega/ck)H_0\,e^{-kz-i\omega
t}$, so that the surface impedance, see e.g. \cite{LL1}:
\begin{equation}
\zeta= -\frac{E_y}{H_x}\Big |_{z=0}=-\,\frac{i\omega}{ck}\,.
\label{zeta}
\end{equation}

If  $\delta\gg\lambda$,
\begin{equation}
k\approx\frac{1}{\lambda}\left(1-i\,\frac{\lambda^2}{\delta^2}\right)
\label{eq5}
\end{equation}
and
\begin{equation}
\zeta\approx\frac{\omega\lambda^3}{c\delta^2}-i\,\frac{\omega\lambda }{c} \,.
\label{zeta1}
\end{equation}

Thus, the dissipative part of impedance is given by
\begin{equation}
 {\rm Re}\,\zeta \approx\frac{2\pi}{c^3}
\omega^2\sigma\lambda^3\,.
\label{imp}
\end{equation}
 The imaginary part of the impedance  is not
affected by quasiparticles part of the current, see e.g.
Ref.\,\onlinecite{LL1}, i.e. it depends only on $\lambda$.
It is worth noting that Eqs.\,(\ref{eq5}) and (\ref{zeta1}) do not hold in immediate vicinity of $T_c$, where $\lambda$ diverges.

%%%%%%
 \subsubsection{Susceptibility of a slab}
 %%%%%%

 It is instructive to consider   Eq.\,(\ref{London}) for a superconducting slab   of the thickness $d$ in the applied ac field $H_x=H_0e^{-i\omega t}$ parallel to the slab faces. The solution is
  \begin{eqnarray}
  H_x=H_0\frac{\cosh(kz)}{\cosh(kd/2)}e^{-i\omega t} \,.
\label{H-slab}
\end{eqnarray}
  with $k$ of Eq.\,(\ref{k}) and $z$ counted from the slab middle. The electric field is
   \begin{eqnarray}
 E_y=i\frac{\omega H_0}{ck}\,\frac{\sinh(kz)}{\cosh(kd/2)}e^{-i\omega t} \,,
\label{E-slab}
\end{eqnarray}
   and the surface impedance
 \begin{equation}
\zeta= -\frac{E_y}{H_x}\Big |_{z=d/2}=- \frac{i\omega}{ck}\tanh\frac{kd}{2}\,.
\label{zeta-slab}
\end{equation}

Commonly measured quantity is the susceptibility defined as ratio of the average magnetization   $\mu$ of  the slab   to the applied field:
 \begin{eqnarray}
\chi&=&  \frac{\mu_x}{H_0}=\frac{1}{4\pi d}\int_{-d/2}^{d/2}\frac{H_x(z)-H_0}{H_0}dz\nonumber\\
&=& -\frac{1}{4\pi} +\frac{1}{2\pi dk}\tanh\frac{kd}{2}\,.
\label{chi-slab}
\end{eqnarray}

Hence, we have a simple relation between the surface impedance and the slab susceptibility:
  \begin{eqnarray}
\chi +\frac{1}{4\pi} =\frac{ic}{2\pi d\omega}\,\zeta\,,
\label{zeta-chi}
\end{eqnarray}
i.e. the surface impedance is proportional to the deviation of susceptibility from the  Meissner value $-1/4\pi$.

Hence, for $\lambda\ll\delta$ one obtains with the help of Eq.\,(\ref{zeta1}):
  \begin{eqnarray}
\chi +\frac{1}{4\pi} =\frac{\lambda}{2\pi d}+i\frac{\lambda^3}{2\pi d\delta^2}\,\,.
\label{thick_slab}
\end{eqnarray}

%%%%%%
 \section{Anisotropic materials}
 %%%%%%

 In the absence of vortices, the order parameter can be taken as  real  so that     the current equation  becomes
 \begin{equation}
 J_k= \sigma_{kl}  E_l -\frac{c}{4\pi}\left(\lambda^{-2}\right)_{kl}\,  A_l  \,,
\label{current1}
\end{equation}
where $\sigma_{kl}$ and $\left(\lambda^{-2}\right)_{kl}$ are tensors of the conductivity due to normal excitations and of the inverse square of the penetration depth. As usual,  summation is implied over double indices. Being interested in problems with no conversion of  normal currents to super-currents, we impose the conditions
 \begin{equation}
 {\rm div}\bm J_n= \sigma_{kl} \frac{\partial E_l}{\partial x_k}=0\,,\quad   {\rm div}\bm J_s= \lambda^{-2} _{kl}\, \frac{\partial A_l}{\partial x_k}= 0\,,
\label{div=0}
\end{equation}
i.e. the densities of   normal excitations and of Cooper pairs are separately conserved.
In particular, this implies a certain gauge for the vector potential.

 In order to obtain an equation for magnetic field exclusively, one has to isolate $\bm E$ and apply  the Maxwell equation curl$\bm E =-\partial_t\bm H/c$. To this end,  multiply  Eq.\,(\ref{current1}) by $\sigma_{sk}^{-1}=\rho_{sk}$ with $\rho_{sk}$ being the resistivity tensor and sum up over $k$:
 \begin{equation}
\rho_{sk} J_k=   E_s -\frac{c}{4\pi}\rho_{sk}\lambda^{-2} _{kl}\,  A_l  \,.
\label{current2}
\end{equation}

In the following it is convenient to use the notation curl$_u\bm E= \epsilon_{uvs}\partial E_s/\partial x_v$ where $\epsilon_{uvs}$ is Levi-Chivita unit antisymmetric tensor:     all components with even number of transpositions from $(xyz)$ are +1, $-1$ for the odd ones, and zero otherwise. Hence, applying $\epsilon_{uvs}\partial /\partial x_v$ to Eq.\,(\ref{current2}), one obtains anisotropic London equations for  the magnetic field, the main result of this paper:
 \begin{eqnarray}
\frac{c}{4\pi} \rho_{sk} \epsilon_{uvs} \epsilon_{kmn}\frac{\partial^2H_n}{\partial x_v\partial x_m}& +&\frac{\partial H_u}{c\,\partial t}  \nonumber\\
 &  =&    -\frac{c}{4\pi}\rho_{sk}\lambda^{-2} _{kl}\,\epsilon_{uvs} \frac{\partial A_l}{\partial x_v}  \,.\qquad
\label{anis-London}
\end{eqnarray}

One can  check    that in the isotropic case this equation reduces to the time-dependent London equation $(\ref{London})$.
Another limit to check is the static anisotropic London equations \cite{K81}. In this case we have
 \begin{eqnarray}
 \rho_{sk} \epsilon_{uvs}\left( \epsilon_{kmn}\frac{\partial^2H_n}{\partial x_v\partial x_m} +   \lambda^{-2} _{kl}\,  \frac{\partial A_l}{\partial x_v}\right) =0 \,.\qquad
\label{anis-static}
\end{eqnarray}
Clearly, this equation is satisfied if
\begin{eqnarray}
  \epsilon_{kmn}\frac{\partial H_n}{ \partial x_m} +   \lambda^{-2} _{kl}\,    A_l   =0 \,.
\label{anis-static1}
\end{eqnarray}
We now introduce a tensor $\left(\lambda^2\right)_{kl}$ inverse to $\left(\lambda^{-2}\right)_{kl}$, multiply the last equation by $\left(\lambda^2\right)_{k\mu}$, and sum up over $k$:
\begin{eqnarray}
\lambda^2_{k\mu}  \epsilon_{kmn}\frac{\partial H_n}{ \partial x_m} +     A_\mu   =0 \,.
\label{anis-static2}
\end{eqnarray}
Finally,  apply to this   $\epsilon_{u v\mu}\partial /\partial x_v$ to replace curl$\bm A$ with $\bm H$  and  obtain static anisotropic London equations  \cite{K81}.

%%%%%%
 \subsection{Orthorhombic slab with plane faces ${\bm a}{\bm b}$  }
 %%%%%%

 Cumbersome Eqs.\,(\ref{anis-London}) are   applicable in  coordinate system $(x,y,z)$ oriented arbitrarily relative to the anisotropic sample.
 One, of course, can choose  $(x,y,z)$ as the crystal  frame $(a,b,c)$ where     $ \rho_{sk}$ and $ \lambda^{-2} _{kl}$  are diagonal.
  Consider a slab of  a thickness $d$ of orthorhombic material with $a,b$ (or  $x,y $) plane faces;    $z$ is counted from the slab middle.
Let the ac applied field $\bm H_0$ be parallel to $x$; the field inside the slab depends only on $z$.

Consider the first term in Eq.\,(\ref{anis-London}).
Since $v$ and $m$ take only $z$ values and $n=x$, it is readily seen that this term reduces to
 \begin{eqnarray}
-\frac{c}{4\pi} \rho_{yy}  \frac{\partial^2H_x}{\partial z^2 }  \,.
\label{1st}
\end{eqnarray}
 The term on the right of Eq.\,(\ref{anis-London}) can be treated similarly to obtain  $c\rho_{yy}\lambda_{yy}^{-2}\partial_z A_y$. Hence
  \begin{eqnarray}
-   \frac{\partial^2H_x}{\partial z^2 } +\lambda_{yy}^{-2}H_x + \frac{4\pi}{c^2\rho_{yy}} \,\frac{\partial H_x}{\partial t}=0.
\label{TDLx}
\end{eqnarray}
This equation is equivalent to the isotropic Eq.\,(\ref{LondonHx}) with the same solution (\ref{H-slab}) for the slab, but now
   \begin{eqnarray}
k_x^2=  \lambda_{yy}^{-2}  - \frac{2i}{\delta^2_{yy}} \,,\qquad \delta^2_{yy} = \frac{c^2}{2\pi \sigma_{yy}\omega}.
\label{k_x2}
\end{eqnarray}
As expected, the decaying behavior of   $H_x$ is determined by characteristics of persistent and normal currents in the $y$ direction.

Thus, the isotropic result for the susceptibility is directly translated to this situation. In particular,
  if $\lambda_{yy}\ll\delta_{yy}$ one obtains for the component $\chi_{xx}$ of the susceptibility tensor:
  \begin{eqnarray}
\chi_{xx} +\frac{1}{4\pi} =\frac{\lambda_{yy}}{2\pi d}+i\frac{\lambda_{yy}^3}{2\pi d\delta_{yy}^2}\, .
\label{chi_xx}
\end{eqnarray}
If the applied field is directed along $y$, the same  argument leads to:
  \begin{eqnarray}
\chi_{yy} +\frac{1}{4\pi} =\frac{\lambda_{xx}}{2\pi d}+i\frac{\lambda_{xx}^3}{2\pi d\delta_{xx}^2}\, .
\label{chi_yy}
\end{eqnarray}
These formulas cannot be used  too close to $T_c$ where the inequality $\lambda\ll\delta$ is violated.

It is worth noting that the anisotropy of the penetration depth is related to the anisotropy of susceptibility:
    \begin{eqnarray}
\gamma_\lambda = \frac{\lambda_{xx} }{\lambda_{yy} } \approx \frac{{\rm Re}\,\chi_{yy} +1/4\pi}{{\rm Re}\,\chi_{xx} +1/4\pi}\,.
    \label{ratio1}
\end{eqnarray}
 Taking the ratio of imaginary parts we obtain
    \begin{eqnarray}
\frac{{\rm Im}\,\chi_{yy}  }{{\rm Im}\,\chi_{xx}  }\approx  \frac{ \delta_{yy}^2}{\delta_{xx}^2} \frac{\lambda_{xx}^3}{\lambda_{yy}^3} = \gamma_\sigma\gamma_\lambda^3 \, ,
    \label{ratio3}
\end{eqnarray}
where $ \gamma_\sigma =\sigma_{xx}/\sigma_{yy}$.

\subsubsection{Angular dependence of susceptibility}

Let the applied field at the sample surface be   at an angle $\varphi$ with the $a$ axis, $\bm H=H_0 \left(\hat{\bm x} \cos\varphi   +\hat{\bm y} \sin\varphi  \right)$.  Since London and Maxwell equations are linear, the solution   is the superposition of two solutions for applied fields oriented along the principle directions:
   \begin{eqnarray}
\bm H=H_0 \left[\hat{\bm x}\frac{\cos\varphi \cosh(k_xz)}{\cosh(k_x d/2)} +\hat{\bm y}\frac{\sin\varphi \cosh(k_yz)}{\cosh(k_y d/2)}\right] \qquad
\label{h_appl_phi}
\end{eqnarray}
where the factor $e^{-i\omega t}$ is omitted for brevity. It is worth noting that since the decay lengths  for the magnetic field along $\hat{\bm x}$ (on the order of $1/k_x$) differs from $1/k_y$, the field rotates with increasing depth $z$.
In this situation, the magnetic moment $\bm \mu$ will have not only the component parallel to the applied field, $\mu_\parallel$, but  a perpendicular component  as well.

One has for the electric field:
    \begin{eqnarray}
\bm E  = \frac{i\omega H_0}{c}\left[\frac{\hat{\bm x}\, \sin\varphi \sinh(k_yz)}{k_y\cosh(k_y d/2)}   -
\frac{\hat{\bm y}\, cos\varphi \sinh(k_xz)}{k_x\cosh(k_x d/2)}     \right]. \qquad
\label{el_field_phi}
\end{eqnarray}

The commonly measured susceptibility    is  defined as
   \begin{eqnarray}
\chi_\parallel= \frac{\mu_\parallel }{H_0}&=& \frac{ \mu_x \cos\varphi + \mu_y \sin\varphi}{H_0} \nonumber\\
&=& \chi_{xx}\cos^2\varphi +\chi_{yy}\sin^2\varphi .
\label{mu_par}
\end{eqnarray}
where $\chi_{xx}$ and $\chi_{yy}$ are given in Eqs\,(\ref{chi_xx}) and (\ref{chi_yy}).  This gives:
   \begin{eqnarray}
{\rm Re}\chi_\parallel &=&-\frac{1}{4\pi} +\frac{1}{2\pi d} \left(\lambda_{yy}\cos^2\varphi +\lambda_{xx}\sin^2\varphi
\right) , \nonumber\\
{\rm Im}\chi_\parallel &=& \frac{1}{2\pi d} \left(\frac{\lambda_{yy}^3}{\delta_{yy}^2}\cos^2\varphi    +\frac{\lambda_{xx}^3}{\delta_{xx}^2}  \sin^2\varphi  \right).
\label{ReIm}
\end{eqnarray}

\subsubsection{Dissipation and torque}

Given the fields at the surface $z=\pm d/2$, one evaluates the Pointing vector, i.e. the energy flux into the sample and the dissipation power  \cite{LL1}. One obtains after straightforward algebra:
    \begin{eqnarray}
\overline{S}_z&=&-\frac{c}{8\pi}{\rm Re} (\bm E\times  \bm  H_0^* )_{z=d/2}\nonumber\\
&=&  \frac{\omega H_0^2}{8\pi}  \frac{ \lambda_{xx}^3}{\delta_{xx}^2}\left[  \sin^2\varphi+\left(\frac{ \lambda_{yy}}{\lambda{xx}}\right)^3  \left(\frac{ \delta_{xx}}{\delta{yy}}\right)^2\cos\varphi^2\right]. \qquad
\label{S}
\end{eqnarray}
Here, $\overline{S}_z$ denotes the time average over the period $2\pi/\omega$.
If the parameter
    \begin{eqnarray}
p= \left(\frac{ \lambda_{yy}}{\lambda{xx}}\right)^3  \left(\frac{ \delta_{xx}}{\delta{yy}}\right)^2  >1\,,
\label{parameter}
\end{eqnarray}
  $\cos^2\varphi$ dominates and the dissipation has minimum at $\varphi=\pi/2$, i.e, for $\bm H_0$ directed along $y$. If   $p<1$, the dissipation is minimal for the field ${\bm H}_0$ directed along $x$. Since the system prefers the state  with minimum dissipation, one expects a torque for $0<\varphi<\pi/2$ acting to rotate the sample to this state.

This conclusion is   confirmed by calculating the torque $\bm\tau$ averaged over the AC period:
    \begin{eqnarray}
\overline{\tau_z}=\frac{1}{2} {\rm Re}(\bm \mu \times \bm H_0^*)=\frac{1}{2} {\rm Re}( \mu_x   H_{0y}^* -\mu_y   H_{0x}^* )
\label{torque}
\end{eqnarray}
where $\mu_x=\chi_{xx}H_0\cos\varphi$ and $\mu_y=\chi_{yy}H_0\sin\varphi$.
We obtain:
    \begin{eqnarray}
\overline{\tau_z}=\frac{H_0^2}{8\pi d} (\lambda_{yy}-\lambda_{xx})\sin 2\varphi .
\label{torque1}
\end{eqnarray}

\section{Discussion}

Anisotropic London equations taking into account normal currents are derived and applied for evaluation of the surface impedance and susceptibility $\chi$ for a simple geometry in which sample surfaces coincide with the $ab$ planes of orthorhombic crystal. In principle, applying the ac field along $a$ and $b$ crystal axes one can extract both  $\chi_{aa}$ and $\chi_{bb}$ of the susceptibility tensor.

In usual situation of the penetration depth small relative to the skin depth, the deviation of real part of susceptibility from Meissner's $-1/4\pi$ depends only on $\lambda$, so that the ratio of these deviations for two principle directions gives the anisotropy parameter $\gamma_\lambda=\lambda_{aa}/\lambda_{bb}$, Eq.\,(\ref{ratio1}). Hence, $\gamma_\lambda $ can, in principle,  be extracted from microwave susceptibility data.
The behavior of $\gamma_\lambda $ with temperature is of intense interests in studies of new materials and it remains to be seen whether or not experimental complications related to finite size of actual samples can be overcome \cite{Pro}.

Given the slab geometry we consider in this paper, thick anisotropic films seems the best  to check our formulas. Strongly anisotropic properties of cuprates makes them good candidates for such measurements. One can find plenty of information for these possibilities in Ref.\,\onlinecite{Narlikar}.

While deep in the superconducting state the contribution of normal quasiparticles to susceptibility is much smaller than the  Meissner contribution by a factor  $\sim \lambda^2/\delta^2$, see Eq.\,(\ref{thick_slab}), only the latter is frequency-dependent via the skin depth $\delta (\omega)$. Therefore, one can measure the response as a function of  $\omega$  and extract the  $\omega$-dependent part. In fact, Eq.\,(\ref{thick_slab}) can be written as $\chi +1/4 \pi = A+iB \omega$ with $\omega$-independent $A,B$. Therefore the derivative of the response with respect to frequency will provide  the imaginary part of $\chi$.

Another quantity which can  be extracted from the susceptibility data is the conductivity of normal excitations $\sigma$. It coincides with the normal state conductivity near $T_c$ (for gapless superconductors, for all temperatures). However, experimentally,  little is known about this conductivity away of $T_c$. Still, this quantity is of interest, in particular, given recent theoretical work of Smith, Andreev, and Spivak stating that the conductivity can be strongly enhanced due to inelastic scattering \cite{Andreev}. The anisotropy of $\sigma$ can, in principle, be extracted from the ratio of imaginary parts of susceptibility and the anisotropy of the penetration depth, Eq.\,(\ref{ratio3}).

It should be noted that we applied the general Eqs.\,(\ref{anis-London}) to an   infinite slab. In experiments, one deals with finite samples. In this case, magnetic susceptibility measured in the applied field along, say $b$-axis in addition to $\lambda_{xx} $  will also depend on $\lambda_{zz} $.
What is worse, the sample shape will give an extra angular modulation   when the angle $\varphi$ of the applied field direction is swept.
These and other difficulties which may arise in measurements of the susceptibility of anisotropic samples and possible ways to overcome them are discussed elsewhere \cite{Pro}.

\section{Acknowledgement}

%I am grateful to Ruslan Prozorov for many helpful discussions.
 This work was supported by the U.S. Department of Energy (DOE), Office of Science, Basic Energy Sciences, Materials Science and Engineering Division.  Ames Laboratory  is operated for the U.S. DOE by Iowa State University under contract \# DE-AC02-07CH11358.

\end{document}